\documentstyle[12pt,aaspp4]{article}

\begin{document}

\title {Near Infrared Photometry of Galactic Globular Clusters $\rm M\,56$ 
and $\rm M\,15$. Extending the Red Giant Branch vs. Metallicity Calibration 
Towards Metal Poor Systems.
\footnote{Based on data taken at the Steward Observatory 2.3m Bok 
Telescope equipped with the 256x256 near IR camera.}
}

\author{Valentin D. Ivanov}
\affil{Steward Observatory, The University of Arizona, 933 N. Cherry 
Avenue, Tucson, AZ 85721; vdivanov@as.arizona.edu}

\author{Jordanka Borissova}
\affil{Institute of Astronomy, Bulgarian Academy of Sciences, 
72~Tsarigradsko chauss\`ee, BG\,--\,1784 Sofia, Bulgaria, 
jura@haemimont.bg} 

\author{Almudena Alonso-Herrero}
\affil{Steward Observatory, The University of Arizona, 933 N. Cherry 
Avenue, Tucson, AZ 85721; aalonso@as.arizona.edu}

\and

\author{Tatiana Russeva}
\affil{Institute of Astronomy, Bulgarian Academy of Sciences, 
72~Tsarigradsko chauss\`ee, BG\,--\,1784 Sofia, Bulgaria}

\begin{abstract}
Infrared $\rm JK_{\rm s}$-band photometry of the Galactic globular 
clusters $\rm M\,15$ and, for the first time, $\rm M\,56$ is presented. 
We estimated the reddening ($\rm E(B-V)=0.18\pm0.08$ mag) and distance 
modulus ($\rm ({\it m}-M)_V=15.43\pm0.30$ mag) towards the poorly 
studied globular cluster $\rm M\,56$. We combined our data with 
observations of other clusters from the literature (12 in total) to 
extend the ${\rm [Fe/H]}$ vs. Red Giant Branch (RGB) slope relation 
towards metal-poor clusters. Our best fit yields to 
${\rm [Fe/H]}\,=\,-3.40(\pm0.22)\,-27.74(\pm2.35)\,\times\,(RGB\,Slope)$,  
with an $\rm r.m.s.\,=\,0.20$. The broader metallicity baseline 
greatly reduced the uncertainties compared to other existing 
calibrations. We confirmed a previously obtained calibration of 
the relation between the RGB color $\rm (J-K_{\rm s})_0(RGB)$ at 
$\rm M_{K_{\rm s}}=-5.5$ vs. $\rm[ Fe/H]$: 
${\rm [Fe/H]}\,=\,-6.90(\pm0.99)\,+6.63(\pm1.05)\,\times\,(J-K_{\rm s})_0(RGB)$
with an $\rm r.m.s.\,=\,0.33$. 
Finally, using the new RGB slope calibration we estimated the abundance 
of the super metal-rich cluster Liller 1 ${\rm [Fe/H]}\,=\,0.34\pm0.22$. 
\end{abstract}

\keywords{Stars: Population II -- Stars:
	Hertzsprung-Russell (HR) diagram -- Galaxy: globular
	clusters: individual: $\rm M\,15$ -- Galaxy: globular
	clusters: individual: $\rm M\,56$ -- Galaxy: globular
	clusters: individual: $\rm NGC\,7099$ -- Galaxy: globular
	clusters: individual: $\rm NGC\,6553$ -- Galaxy: globular
	clusters: individual: Liller 1}

\section{Introduction}

Globular clusters are a fundamental laboratory for the study and 
understanding of stars and their evolution. They offer a unique 
opportunity to study samples of stars with a single age and 
metallicity, and provide a sequence of parameters which describe 
the stellar populations as ensembles of stars. Those parameters 
include the position of the red giant branch (RGB), the relative 
populations of the blue and red horizontal branches, and the 
relative number of RR Lyr variables. 

Davidge et al. (1992\markcite{dav92}) considered for the first time 
the slope of the RGB on a combined optical-infrared color-magnitude 
diagram (CMD) as a metallicity indicator. This technique is 
particularly promising (Kuchinski, Frogel \& Trendrup 
1995\markcite{kuc95}; Kuchinski \& Frogel 1995\markcite{kuc95b}; 
Ferraro et al., 2000\markcite{fer00}), because it is not affected by 
the reddening towards the cluster. Bypassing the reddening correction 
can help to improve significantly the photometric determination of 
the metallicity of more distant and highly obscured systems. 

The stars on the RGB radiate most of their energy in the infrared, 
with the added advantage that the reddening is greatly diminished 
in this part of the spectrum. Although both those arguments are not 
relevant for observations of Galactic globular clusters, they become 
critical in the studies of distant systems, where the patchy internal 
extinction is added to the foreground extinction in the Milky Way, 
which makes the interpretation of the data more difficult. Also, 
many potential targets of interest (such as nearby dwarf galaxies) 
are expected to have a low metal content. Until recently, the 
available calibrations did not span all the necessary range of 
$\rm[Fe/H]$. Indeed, Kuchinski \& Frogel (1995\markcite{kuc95b}) 
based their calibration on a set of Galactic globular clusters with 
metallicities from $\rm[Fe/H]$~=~1.01 to -0.25. Ferraro et al. 
(2000\markcite{fer00}) presented for the first time high quality 
near infrared CMDs of 10 Galactic globular clusters, and a detailed 
analysis of the RGB behavior as a function of metallicity. 

The goal of this work is to increase the statistical basis of the 
RGB slope vs. $\rm[Fe/H]$ calibration, and to test the existing 
calibrations (Kuchinski, Frogel \& Trendrup 1995\markcite{kuc95}; 
Kuchinski \& Frogel 1995\markcite{kuc95b}; Ferraro et al., 
2000\markcite{fer00}). Clusters with low metallicities are of 
particular interest, because they will allow to make this tool 
applicable to metal-poor stellar systems. We will also increase the 
statistical basis of the RGB slope vs. $\rm[Fe/H]$ calibration. 

We present here $\rm JK_s$ photometry of the central area of 
$\rm M\,15$ and, for the first time, $\rm M\,56$. The basic data for 
the clusters are given in Table~\ref{tbl-1} (Harris 
1996\markcite{har96}; June 22, 1999 version). They are both extremely 
metal-poor clusters. The table contains data on two more clusters 
($\rm M\,30$ and $\rm NGC\,6553$) which we collected from the 
literature. 

\placetable{tbl-1}

$\rm M\,15$ is a well studied Galactic globular cluster. It possesses 
one of the highest known central densities (Yanny et al. 
1994\markcite{yan94}). King (1975\markcite{kin75}) and Bahcall \& 
Ostriker (1975\markcite{bah75}) speculated that the cluster might have 
undergone a core collapse or might contain a central black hole. Sandage 
(1970\markcite{san70}) obtained ground-based photometry of the outer 
region of $\rm M\,51$ and detected an extended blue horizontal branch 
(hereafter HB) but no significant population of blue stragglers. 
Subsequent photometric works were presented  by Auri\`{e}re \& Cordoni 
(1981\markcite{aur81}), Buonano et al. (1985\markcite{buo85}), Bailyn et 
al. (1988\markcite{bai88}), and Cederbloom et al. (1992\markcite{ced92}). 
More recently, the cluster was observed in the optical with the {\it HST} 
by Ferraro \& Parsce (1993\markcite{fer93}) and Yanny et al. 
(1994\markcite{yan94}). Frogel, Persson \& Cohen (1983\markcite{fro83}) 
published $\rm JHK$ measurements of five bright red giants in 
$\rm M\,15$. The most recent variability study of $\rm M\,15$ (Buter et 
al., 1998\markcite{but98}) reported light curves of 30 confirmed 
variable stars, mostly RR Lyr. 

In contrast, $\rm M\,56$ is surprisingly poorly studied, probably because 
is lays close to the Galactic plane $(l=62.66\arcdeg, b=+8.34\arcdeg)$. 
Rosino (\markcite{ros51}1951) obtained the first photographic CMD of this 
cluster. Barbon (1965\markcite{bar65}) built the first CMD in the 
standard $BV$ colors. Smriglio, Dasgupta \& Boyle (1995\markcite{smr95}) 
used the Vilnius photometric system to estimate the extinction towards 
$\rm M\,56$, and pointed to a possible reddening variation across the 
cluster area. A number of observations of variable stars in $\rm M\,56$ 
have been undertaken throughout the years (Sawyer 1940\markcite{saw40}; 
Sawyer 1949\markcite{saw49}; Rosino 1961\markcite{ros61}; Wehlau \& 
Sawyer Hogg 1985\markcite{weh85}). The latest CMD for this cluster 
(Grundahl et al. 1999\markcite{gru99}) is in the Str\"{o}mgren $(u,u-y)$ 
system and shows very well defined blue and red HBs.

\section{Observations and Data Reduction}

We obtained $\rm JK_{\rm s}$ imaging of $\rm M\,56$ and $\rm M\,15$ 
using a $256\times256$ NICMOS3 array at the 2.3-m Bok Telescope of the 
University of Arizona on Kitt Peak, with a plate scale of $0.6\,$arcsec 
pixel$^{-1}$, under photometric conditions on Nov 5, 1998. The average 
seeing during the observations was 1.0-1.2 arcsec. The observational 
strategy consisted of taking cluster images interleaved with sky images 
$6\arcmin-7\arcmin$ away from the targets. We dithered both object and 
sky images to improve the bad pixel and cosmic ray corrections. 

The data reduction included subtraction of dark current frames, 
flat-fielding with median combined empty sky frames, and sky 
subtraction using IRAF. 
\footnote{IRAF is distributed by the National Optical Astronomy 
Observatories, which are operated by the Association of Universities 
for Research in Astronomy, Inc., under cooperative agreement with 
the National Science Foundation.}
The images were shifted to a common position with cubic spline 
interpolation, and averaged together to produce the final image. The 
photometric calibration was performed using observations of standard 
stars from the list of \markcite{eli82}Elias et al. (1982). Although 
we used the $\rm K_{\rm s}$ filter which has shorter longer wavelength 
transmission limit than the $\rm K$ standard filter, the two photometric 
system are nearly identical (Persson et al. 1998\markcite{per98}) 
within the observational errors. The photometric calibration errors 
associated with the standard stars scatter are 0.05, and 0.06 mag in 
$\rm J$, and $\rm K_{\rm s}$ respectively. 

The stellar photometry of the final combined frames was carried out 
using DAOPHOT II (\markcite{ste93}Stetson 1993). We found some small 
variations in the FWHM of the PSF $(\approx 0.08$ arcsec) between the 
inner and outer frame regions. A variable PSF was constructed using a 
large number of moderately bright, isolated stars. We assumed that 
the PSF varied linearly with the position in the frame. A subset of 
the photometric data is presented in Table~\ref{tbl-2} where the 
coordinates are given in pixels relative to the cluster centers, and 
the last two columns contain the formal DAOPHOT errors.  

\placetable{tbl-2}

The formal DAOPHOT errors shown in Figure~\ref{fig-1} demonstrate the 
internal accuracy of the photometry. The typical errors down to 
$\rm J\leq16$ and $\rm K_{\rm s}\leq16$ are smaller than 0.1 mag. The 
larger spread of errors in $\rm M\,15$ is due to the denser central 
core of this cluster. To account for the uncertainty of the sky 
subtraction we added in quadrature 0.01 mag in $\rm J$, and 0.02 mag 
$\rm K_{\rm s}$ to these errors. 
To estimate independently the internal accuracy of our photometry we 
carried out an artificial star simulation. This is the most complete 
technique for error determination because it includes the sky 
background variations, crowding errors, and the PSF variations across 
the field. We added 100 artificial stars with known brightnesses at 
random places on the $\rm J$ and $\rm K_{\rm s}$ images of each cluster. 
We measured then their magnitudes in the same manner as for the program 
stars. We repeated this simulation ten times and calculated the mean 
standard deviations for given magnitude bins (Table~\ref{tbl-3}). We 
successfully recovered the formal DAOPHOT errors (Figure~\ref{fig-1}). 
The former errors are small compared with the photometric calibration 
errors, and thus we used the DAOPHOT errors throughout the paper 
taking advantage of the individual error estimates for each star. 

\placefigure{fig-1}
\placetable{tbl-3}

We have one star in common with Frogel, Cohen \& Persson 
(1983\markcite{fro83}) - I-12 in their notation. It is at the edge of 
our field. They estimated $\rm K=9.42$, and $\rm J=10.19$. Our 
measurements are $\rm K_{\rm s}=9.37\pm0.06$, and $\rm J=10.29\pm0.06$ 
where the formal DAOPHOT errors and the photometric calibration errors 
are added in quadrature. The corresponding differences are 0.05 mag and 
0.10 mag, acceptable if compared with the errors. Undoubtedly some of 
the problem in $\rm K_{\rm s}$ may arise from the different photometric 
systems. Persson et al. (1998\markcite{per98}; see their Table 3) showed 
that for red stars $\rm K_{\rm s}$ and $\rm K$ are rarely further apart 
than 0.02 mag.

\section{Color-Magnitude Diagrams}

The $\rm K_{\rm s}$, $\rm J-K_{\rm s}$ CMD for $\rm M\,15$ and 
$\rm M\,56$ datasets are presented in Figure~\ref{fig-2}. Only stars 
with DAOPHOT errors of less than 0.06 for $\rm K_{\rm s}\leq14.0$~mag, 
(filled circles) and stars with errors less than 0.10 for 
$\rm K_{\rm s}\geq14.0$ (open circles) were included. To minimize the 
field star contamination in $\rm M\,56$ for stars with $\rm K_{\rm s}$ 
brighter than $14.0$, only stars within the radius $r\,=\,1.16\arcmin$ 
(\markcite{har96}Harris 1996) were included. The stars from within 7 
times the core radius of $\rm M\,15$ ($r_{core}=0.07\arcmin$, 
\markcite{har96}Harris 1996) were excised.

\placefigure{fig-2}

\subsection{$\rm M\,15$}

The giant branch of $\rm M\,15$ is very well defined up to 
$\rm K_{\rm s}=9.5$ mag. The position of the brightest non-variable 
star suggests that the RGB tip lies at $\rm (J-K_{\rm s})=0.92$ mag 
and $\rm K_{\rm s}=9.37$ mag. None of the red variables listed in 
Clement (\markcite{cle99}1999) lies in our field. 

The HB can be identified at $\rm K_{\rm s}=14.35\pm0.3$ mag, derived 
as an average of 25 RR Lyr stars (represented by diamonds in 
Figure~\ref{fig-2}; Clement \markcite{cle99}1999). Unfortunately our 
observations do not span long enough time interval to calculate the 
average K-band magnitude of each RR Lyr star. Instead, the plotted 
RR Lyr magnitudes represent their snapshot brightnesses at the moment 
of the observation. The typical amplitude of RR Lyr in the infrared 
is 0.2-0.3 mag (Carney et al., 1995\markcite{car95}). Combined with 
the average photometric error (0.17 mag for K=14-16), it accounts for 
the HB uncertainty. 

The red HB spans a range from $\rm (J-K_{\rm s})=0.46$ to $0.35$ mag. 
On the $\rm (V,B-V)$ CMD, $\rm M\,15$ shows the typical HB morphology 
of metal-poor clusters, with a high blue-to-red HB star ratio, and 
large number of RR Lyr variables (\markcite{dur93}Durrell \& Harris 
1993). Our data are not deep enough to detect the blue HB stars.

\subsection {$\rm M\,56$}

The giant branch is well defined up to $\rm K_{\rm s}\approx10$ mag. 
The RGB tip lies at $\rm (J-K_{\rm s})=0.94$ mag and 
$\rm K_{\rm s}=9.72$ mag. The horizontal branch can be identified at 
$\rm K_{\rm s}=14.45\pm0.02$ mag. The clump at $\rm (J-K_{\rm s})=0.45$ 
mag constitutes the red HB, and the stars with $\rm (J-K_{\rm s})$ 
between 0.10 and 0.20 mag and $\rm K_{\rm s}$ between 14.45 and 15.60 
mag are the blue HB. Unfortunately our photometry is not complete at 
this level to determine the blue-to-red HB star ratio. Since the 
cluster is close to the Galactic plane, some background contamination 
would be present. The stars with $\rm (J-K_{\rm s})\approx0.2$ and 
$\rm K_{\rm s}\leq14.5$ mag are a clear example. Among the fainter 
stars we have a mixture of field and member stars, and we do not 
include those in our considerations. 

Clement (\markcite{cle99}1999) lists twelve variables in $\rm M\,56$. 
Our imaging includes only V2 and V6 (marked in Figure~\ref{fig-2}). 
Their membership is confirmed by relative proper motion measurements 
(\markcite{rus81}Rishel, Sanders, \& Schroder 1981). Wehlau \& Sawyer 
Hogg (\markcite{weh85}1985) classified V2 as an irregular red variable 
with small V amplitude. Its position in our CMD confirms it. V6 has a 
well determined 90 day period, and was classified as an RV Tau type 
(Sawyer 1940\markcite{saw40}; Sawyer 1949\markcite{saw49}; Wehlau \& 
Sawyer Hogg 1985\markcite{weh85}). \markcite{rus00}Russeva (2000) 
tentatively identified 7 additional red variable stars, marked in 
Figure~\ref{fig-2} as open diamonds. They all belong to the RGB, and 
are among the brightest and reddest stars in our sample. There are no 
known RR Lyr stars in our field. 

\subsection{Reddening and Distance of $\rm M\,56$}

Our photometry allows to carry out a new determination of the distance 
and reddening to $\rm M\,56$. Since $\rm M\,15$ has a similar metal 
content to $\rm M\,56$, $\rm M\,15$ can be used as a template for the 
intrinsic RGB color. In addition, \markcite{kuc95}Kuchinski \& Frogel 
(1995) demonstrated that the color of RGB at the level of the HB shows 
little or no change with metallicity. 

For $\rm M\,15$ we measured a reddening corrected color at the HB 
level of $\rm (J-K_{\rm s})_{GB,HB,0}=0.58\pm0.04$ mag. The observed 
RGB color of $\rm M\,56$ at the level of HB is 
$\rm (J-K_{\rm s})_{GB,HB}=0.62\pm0.02$ mag. Assuming that the color 
difference is only due to the reddening, we obtained a relative color 
excess of $E(J-K_{\rm s})_{M\,56,M\,15}\,=\,0.04\pm0.04$ mag. Using 
\markcite{rie85}Rieke \& Lebofsky (1985) extinction law we found 
$E(B-V)_{M\,56,M\,15} = 0.08\pm0.08$ mag, and finally a color excess 
of of $\rm M\,56$ is $\rm E(B-V)_{M\,56}=0.18\pm0.08$ mag. The error 
includes the internal photometric error, errors of fiducial lines fits 
of both globular clusters, and the uncertainty of the HB levels. Note, 
that even though the HB level in $\rm M\,15$ has a large uncertainty 
(0.30 mag), this does not affect seriously our reddening estimate 
because of the steep RGB. The calculated reddening of $\rm M\,56$ is 
very close to the value of $\rm E(B-V)=0.20$ mag given by Harris 
(1996\markcite{har96}). 

The comparison of the reddening corrected HB levels of $\rm M\,56$ 
and $\rm M\,15$ can be used to determine the differential distance to 
$\rm M\,56$ with respect to $\rm M\,15$. We obtained 
$\rm K_{s,HB,0}=14.32\pm0.30$ mag, and $\rm K_{s,HB,0}=14.38\pm0.04$ 
mag for $\rm M\,15$ and $\rm M\,56$ respectively, using the reddening 
law from Rieke \& Lebofsky (1985\markcite{rie85}). We adopted for 
$\rm M\,15$ a color excess of $\rm E(B-V)=0.10$ mag and a distance 
modulus of $\rm ({\it m}-M)_{V}=15.37$ mag from \markcite{har86}Harris 
(1996). The distance scale was established by adding to his horizontal 
branch vs. metallicity calibration an empirical evolutionary correction 
to the zero age horizontal branch as determined by Carney et al. (1992\markcite{car92}). The typical distance modulus uncertainty was 
$\pm0.1$ mag. Thus, we find a distance modulus to $\rm M\,56$ of 
$\rm ({\it m}-M)_V=15.43\pm0.30$ mag. Unfortunately the large 
uncertainty in the HB level prevents us from making a better estimate. 
\markcite{har96}Harris (1996) gives $\rm ({\it m}-M)_V=15.65$ mag. If 
corrected for the reddening, it becomes $\rm ({\it m}-M)_0=15.03$ mag. 
This estimate is based on RR Lyr observations by Wehlau \& Sawyer Hogg 
(1985\markcite{weh85}) who obtained $\rm ({\it m}-M)_0=14.81$ mag with 
a different extinction value. Previously Harris \& Racine 
(1979\markcite{har79}) determined 9.7 kpc or $\rm ({\it m}-M)_0=14.81$. 
Our distance estimate is larger than those. We attribute the difference 
to the uncertain HB position in $\rm M\,15$. 

According to Guarnieri et al. (1998\markcite{gua98}), the $\rm K_s$-band 
absolute magnitude of the HB depends on the cluster metallicity: 
$\rm M_{K}(HB)\,=\,-0.2*\rm[Fe/H]\,-1.53$. This relation predicts a 0.06 
mag difference in the HB level of the two clusters, which is smaller 
than the uncertainties in our observed HB positions. It also predicts 
an absolute $\rm K_s$-band magnitude for the HB of $\rm M\,15$ of 
$\rm M_{K}(HB)\,=\,-1.08$ mag, very close to the average 
$\rm <M_{K}(HB)>\,=\,-1.15\pm0.10$ mag value measured by Kuchinski \& 
Frogel (1995b\markcite{kuc95b}) for their sample of globular clusters. 
Our data yield $\rm M_{K}(HB)\,=\,-0.95\pm0.30$, where the error is  
dominated by the spread of the RR Lyr apparent brightnesses.

\section{The Red Giant Branch as a Metallicity Indicator}

\subsection{The Red Giant Branch Slope vs. $\rm [Fe/H]$}

It is well known that both intrinsic colors $\rm (J-K_{\rm s})_0$ and 
$(V-K_{\rm s})_0$ are 
sensitive to metallicity (e.g. \markcite{fro83}Frogel; Cohen \& Persson 
1983, \markcite{bra98}Braun et al. 1998). \markcite{kuc95} Kuchinski et 
al. (1995) and \markcite{kuc95b}Kuchinski \& Frogel (1995) demonstrated 
empirically and theoretically that the slope of the upper RGB in the 
$\rm K_{\rm s}$ vs. $\rm (J-K_{\rm s})$ diagram is sensitive to the 
metallicity for a sample of metal-rich Galactic globular clusters. \markcite{tie97}Tiede, Martini \& Frogel (1997) extended this relation 
to a sample of open clusters and bulge stars. Possible extragalactic 
applications prompted us to extend this calibration towards lower 
metallicity clusters. 

We added to Kuchinski \& Frogel (1995\markcite{kuc95b}) sample our two 
globular clusters $\rm M\,15$, $\rm M\,56$, and two more clusters from 
the literature ($\rm NGC\,7099$ and $\rm NGC\,6553$). $\rm NGC\,7099$ 
is the most metal-poor among the clusters observed in infrared by Cohen 
\& Sleepers (1995\markcite{coh95}), and their photometry includes a 
sufficiently large number of RGB stars. $\rm NGC\,6553$ was studied by 
Guarnieri et al. (1998\markcite{gua98}). Although it is not a 
particularly metal-poor cluster, it will improve the statistical weight 
of our calibration. Gathering data from different sources observed with 
different photometric systems always involves some danger of 
incompatibility. For red stars the average difference between $\rm K$ 
and $\rm K_{\rm s}$ measurements of Persson et al. (1998\markcite{per98}) 
is 0.01 mag with a standard deviation of 0.02 mag. We increased 
correspondingly the uncertainties of the photometry. 

First, we had to separate the giants belonging to the upper RGB. 
Following \markcite{tie97}Tiede, Martini \& Frogel (1997), these are 
stars with absolute $\rm K$ magnitudes spanning the range from $-2$ to 
$-6.5$ mag. However, \markcite{fro88}Frogel \& Elias (1988) and 
Montegriffo et al. (\markcite{mon95}1995) determined that most (if not 
all) of the brightest RGB stars are in fact AGB long-period variables. 
Hence, we excluded from our analysis all the known red variable stars. 
Further, we excluded stars within the central 50 arcsec in both 
$\rm M\,15$ and $\rm M\,56$ to minimize the errors due to field 
crowding. 

As in Kuchinski et al. (1995\markcite{kuc95}) we carried out a 
least-squares fit to the giants from 0.6 mag to 5.2 mag above the HB 
level in $\rm K_{\rm s}$, taking into account errors along both axes. 
This effectively increased the weight of the most accurate measurements, 
which are usually the brightest cluster stars. To exclude possible 
random errors we rejected stars which deviated more than three times the 
r.m.s. of the first fit. This resulted in a loss of only 5, 1, 2 and 4 
stars for $\rm M\,15$, $\rm M\,56$, $\rm NGC\,7077$ and $\rm NGC\,6553$, 
respectively, confirming the good quality of the photometry. The best 
fits for the slopes of the RGB of these clusters are shown in 
Table~\ref{tbl-4}. The errors in this table are purely statistical. The 
fits and the RGB stars used are plotted in Figure~\ref{fig-3}. Varying 
the HB level within the errors, produced an additional 0.002 error in 
the RGB slope, and was added in quadrature to the statistical errors. 
To verify our fitting technique we carried out simulations by creating 
100 artificial RGBs analogous to the $\rm M\,15$ RGB. We started from the 
$\rm K_{\rm s}$-band magnitudes of each star, calculated the corresponding 
$\rm J-K_{\rm s}$, and added random Gaussian errors with the appropriate 
$\sigma$ for each star, along both axes. We were able to recover the 
original RGB slope to within less than $1\sigma$.

\placetable{tbl-4}
\placefigure{fig-3}

The relation of the RGB slope vs. the $\rm[Fe/H]$ is shown in 
Figure~\ref{fig-4}. Fitting a straight line to the data taking into 
account the errors along both axes yields to: 
\begin{equation}
{\rm [Fe/H]}\,=\,-3.40(\pm0.22)\,-27.74(\pm2.35)\,\times\,(RGB\,Slope)
\end{equation}
with the $\rm r.m.s.\,=\,0.20$. 
To test for the compatibility of the data collected from various sources, 
we carried the same fit using only the data from Kuchinski \& Frogel 
(1995\markcite{kuc95b}), and our two clusters. We obtained a fit 
statistically indistinguishable from Equation (1). 

\placefigure{fig-4}

There are three recent determinations of this relation in the literature. 
Kuchinski \& Frogel (1995\markcite{kuc95b}) derived for their sample of 
ten metal-rich clusters: 
\begin{equation}
{\rm [Fe/H]}\,=\,-2.98(\pm0.70)\,-23.84(\pm6.83)\,\times\,(RGB\,Slope)
\end{equation}

Later Tiede, Martini \& Frogel (1997\markcite{tie97}) re-derived it. 
For the sample of twelve globular clusters they obtained:  
\begin{equation}
{\rm [Fe/H]}\,=\,-2.78(\pm0.61)\,-21.96(\pm5.92)\,\times\,(GB\,slope) 
\end{equation}
and if the most metal-poor cluster is rejected: 
\begin{equation}
{\rm [Fe/H]}\,=\,-2.44(\pm0.67)\,-18.84(\pm6.41)\,\times\,(GB\,slope)
\end{equation}

Most recently Ferraro et al. (2000\markcite{fer00}) obtained with much 
higher quality data of 10 clusters: 
\begin{equation}
{\rm [Fe/H]}\,=\,-2.99(\pm0.15)\,-23.56(\pm1.84)\,\times\,(GB\,slope) 
\end{equation} 
They used the metallicity scale of Caretta \& Gratton 
(1997\markcite{car97}). 

Our large range of metallicity improves the fit significantly, although 
the $1\sigma$ errors are slightly larger than those of Ferraro et al. 
(2000\markcite{fer00}). All these derivations are statistically 
indistinguishable from our calibrations. This is particularly important 
when one has to extrapolate the RGB slope vs. $\rm[Fe/H]$ relation 
outside of the explored metallicity range. 

We can now apply our relation to estimate the metallicity of the most 
metal-rich Galactic globular cluster Liller 1. For this cluster 
Armandroff \& Zinn (\markcite{arm88}1988) estimated 
$\rm [Fe/H]=+0.20\pm0.3$ based on integrated light spectroscopy. Frogel, 
Kuchinski \& Tiede (1995\markcite{fro95}) obtained an RGB slope of 
$-0.135\pm0.009$, and derived $\rm [Fe/H]=+0.25\pm0.3$. Using the same 
RGB slope, our fit and that of Ferraro et al. (2000\markcite{fer00}) 
yield metallicities of $\rm [Fe/H]=+0.34\pm0.22$ and 
$\rm [Fe/H]=+0.19\pm0.15$, respectively. These estimates have to be 
taken with caution until more data allow to test whether the RGB slope 
vs. $\rm[Fe/H]$ relation is indeed linear for super metal-rich stellar 
populations.

\subsection{The Red Giant Branch Color vs. $\rm [Fe/H]$}

The slope of the RGB as described here is difficult to measure in 
distant stellar systems because it requires deep photometry, reaching 
the HB level. As somewhat observationally less challenging alternative 
we considered the RGB color at a given absolute magnitude as a 
metallicity indicator. Following Frogel, Cohen \& Persson 
(1983\markcite{fro83}), we chose to calibrate $\rm (J-K_{\rm s})_0(RGB)$ 
at $\rm M_{K_{\rm s}}=-5.5$ mag. 

A drawback of this method is that unlike the RGB slope, the RGB color 
is reddening sensitive and requires a correction prior to the calibration. 
We used the reddening and distance estimates from Harris 
(1996\markcite{har96}). We also assumed that the HB level is always at 
$\rm M_{K_{\rm s}}=-1.15$ mag neglecting the HB luminosity dependence on 
$\rm [Fe/H]$ (Kuchinski \& Frogel 1995\markcite{kuc95b}). 

We calculated the RGB colors from the linear fits to the RGBs 
(Table~\ref{tbl-4}). A test with $\rm M\,15$ and $\rm M\,56$ showed 
that it is identical to averaging the colors across the RGB within the 
observational uncertainties. The small number of stars on the RGB tip 
actually makes the averaging less reliable. In addition, most of the 
RGBs show a high degree of linearity (Kuchinski et al., 
1995\markcite{kuc95}; Kuchinski \& Frogel, 1995\markcite{kuc95b}). 
The errors associated with the RGB colors were calculated as a 
quadrature sum of the r.m.s. of the fit, the errors from the E(B-V) 
$(\approx0.02$ mag) and the errors from the uncertain distance moduli 
$(\approx0.10$ mag). 

Our best fit of $\rm (J-K_{\rm s})_0(RGB)$ at $\rm M_{K_{\rm s}}=-5.5$ vs. 
metallicity (Figure~\ref{fig-5}) was derived with errors along both 
axes: 
\begin{equation}
{\rm [Fe/H]}\,=\,-6.90(\pm0.99)\,+6.63(\pm1.05)\,\times\,({\rm J-K_{\rm s})_0}(RGB)
\end{equation}
with $\rm r.m.s.\,=\,0.33$. We attribute the large r.m.s. to the 
reddening uncertainties.

\placefigure{fig-5}

Frogel, Cohen \& Persson (1983\markcite{fro83}) found: 
\begin{equation}
{\rm [Fe/H]}\,=\,-6.905\,+6.329\,\times\,({\rm J-K_{\rm s})_0}(RGB)
\end{equation}
with r.m.s. = 0.16. This calibration is identical to ours within the 
uncertainties. It was important to confirm their result, because 
although it was based on photometry of a large number of clusters 
(33), it only included 10-20 stars per cluster. 

Minitti, Olszewski \& Rieke (1995\markcite{min95}) also calibrated 
the intrinsic RBG color at $\rm M_{K_{\rm s}}=-5.5$ mag and obtained: 
\begin{equation}
{\rm [Fe/H]}\,=\,-5.0\,+5.60\,\times\,{\rm (J-K_{\rm s})_0}(RGB).
\end{equation}
Their fit is close to ours but shows a slightly different slope. It 
is based on twenty clusters. Unfortunately the severe field star 
contamination towards the Galactic bulge prevented them from a 
reliable estimate of the cluster colors in many cases. 

Ferraro et al. (2000\markcite{fer00}) also performed a similar 
calibration: 
\begin{equation}
{\rm [Fe/H]}\,=\,-4.76(\pm0.23)\,+5.38(\pm0.27)\,\times\,({\rm J-K_{\rm s})_0}(RGB)
\end{equation}
where the fit coefficients are within $3\sigma$ of ours. Obtaining 
better reddening estimates would be crucial for resolving this 
discrepancy.

\section{Summary} 

We showed that the position of the infrared RGB can be used to reliably 
determine the abundance of metal-poor stellar systems with an accuracy 
of $\approx0.2$ (in $\rm [Fe/H]$). Our main results include: 

(1) We presented infrared photometry of the Galactic globular clusters 
$\rm M\,15$ and, for the first time, $\rm M\,56$, and studied the 
morphology of their CMDs. 

(2) We estimated the reddening towards $\rm M\,56$ by comparing the 
RGB color at the level of HB to that of $\rm M\,15$, and obtained 
$\rm E(B-V)=0.18\pm0.08$ mag. We used the relative HB levels of 
the same two clusters to derive a distance modulus to $\rm M\,56$ of 
$\rm ({\it m}-M)_V=15.43\pm0.30$ mag if $\rm ({\it m}-M)_V=15.37$ mag 
is assumed for $\rm M\,15$. 

(3) We compiled a sample of 12 Galactic globular clusters with high 
quality infrared photometry. We recalibrated and extended the RGB 
slope vs. $\rm [Fe/H]$ relation towards low metallicity globular 
clusters. We also reevaluated the RGB color $\rm (J-K_{\rm s})_0(RGB)$ 
at $\rm M_{K_{\rm s}}=-5.5$ mag vs. $\rm[ Fe/H]$ relation. These are 
potentially useful tools to study extragalactic metal-poor stellar 
systems, particularly with high obscuration. Our results independently 
confirm the previously obtained calibrations. 

(3) As an application, we used our newly determined RGB slope vs. 
$\rm [Fe/H]$ relation to estimate the abundance of the super 
metal-rich Galactic globular cluster Liller 1, and obtained 
${\rm [Fe/H]}\,=\,0.34\pm0.22$.

\begin{acknowledgements}
During the course of this work VDI and AA-H were supported by the 
National Aeronautics and Space Administration on grant NAG 5-3042 
through the University of Arizona. The $256 \times 256$ camera 
was supported by NSF Grant AST-9529190. JB and TR were supported by 
the Bulgarian National Science Foundation grant under contract No. 
F-812/1998 with the Bulgarian Ministry of Education and Sciences. 
We are thankful to the anonymous referee for the corrections that 
helped to improve the quality of this paper.
\end{acknowledgements}

\clearpage

\begin{deluxetable}{crrccrcccc}
\tablenum{1}
\tablewidth{0pt}
\tablecaption{Cluster data.\label{tbl-1}}
\small
\tablehead{
\multicolumn{1}{c}{Name}           & \multicolumn{1}{c}{$\rm R_{GC}$} & 
\multicolumn{1}{c}{$\rm R_{\sun}$} & \multicolumn{1}{c}{$\rm ({\it m}-M)_V$} & 
\multicolumn{1}{c}{$\rm E(B-V)$}   & \multicolumn{1}{c}{$\rm S_{RR}$} & 
\multicolumn{1}{c}{$\rm HBR$}      & \multicolumn{1}{c}{$\rm [Fe/H]$} & 
\multicolumn{1}{c}{$\rm r_c$}      & \multicolumn{1}{c}{$\rm r_h$}\nl}
\startdata
$\rm M\,15$/$\rm NGC\,7078$&10.4&10.3&15.37&0.10&24.1&0.67&-2.25&0.07&1.06\nl
$\rm M\,56$/$\rm NGC\,6779$& 9.7&10.1&15.65&0.20& 2.2&0.98&-1.94&0.37&1.16\nl
$\rm M\,30$/$\rm NGC\,7099$& 7.1& 8.0&14.62&0.03&10.7&0.89&-2.12&0.06&1.15\nl
$\rm NGC\,6553$        & 2.5& 5.6&16.05&0.75& 0.6& -  &-0.34&0.55&1.55\nl
\tablecomments{Columns: 
(1) name, 
(2) galactocentric distance in kpc, assuming $R_0=8.0$ kpc, 
(3) distance from the Sun in kpc, 
(4) visual distance modulus, not corrected for extinction, 
(5) Galactic reddening, (6) specific frequency of RR Lyr, 
(7) HB ratio HBR = (B-R)/(B+V+R), 
(8) $\rm[Fe/H]$, 
(9) core radius in arcmin, 
(10) half-mass radius in arcmin}
\tablerefs{\markcite{har96}Harris (1996)}
\enddata
\end{deluxetable}

\begin{deluxetable}{rrrrrrr}
\tablenum{2}
\tablewidth{0pt}
\tablecaption{Photometry of the Galactic globular clusters $\rm M\,15$ 
and $\rm M\,56$.\label{tbl-2}}
\tablehead{
\multicolumn{1}{c}{ID\tablenotemark{a}} & 
\multicolumn{1}{c}{X} 
& \multicolumn{1}{c}{Y} & 
\multicolumn{1}{c}{$\rm K_{\rm s}$} & 
\multicolumn{1}{c}{$\rm J-K_{\rm s}$} & 
\multicolumn{1}{c}{$\rm \sigma(K_{\rm s})$} & 
\multicolumn{1}{c}{$\rm \sigma(J)$}\nl}
\startdata
\multicolumn{7}{l}{$\rm M\,15$}\nl
  1 &  11.64 &  2.21 & 17.32 & 0.45 & 0.37 & 0.36\nl
  2 & 186.90 &  2.34 & 14.38 & 0.48 & 0.08 & 0.06\nl
  3 &  42.45 &  2.72 &  9.37 & 0.92 & 0.03 & 0.01\nl
  4 & 117.50 &  3.15 & 16.80 & 0.39 & 0.17 & 0.15\nl
  5 & 140.80 &  3.31 & 15.99 & 0.23 & 0.10 & 0.10\nl
  6 &  56.02 &  3.50 & 14.53 & 0.60 & 0.05 & 0.04\nl
  7 & 228.20 &  3.60 & 13.63 & 0.53 & 0.04 & 0.02\nl
\multicolumn{7}{c}{...}\nl
\multicolumn{7}{l}{$\rm M\,56$}\nl
  1 & 241.90 & 47.89 & 16.28 & 0.81 & 0.08 & 0.12\nl
  2 & 254.00 & 47.90 & 18.00 & 0.12 & 0.23 & 0.22\nl
  3 &  42.37 & 47.90 & 17.65 & 0.23 & 0.13 & 0.15\nl
  4 &  64.44 & 48.20 & 17.33 & 1.10 & 0.23 & 0.26\nl
  5 & 233.90 & 48.31 & 11.57 & 0.93 & 0.03 & 0.04\nl
  6 & 158.10 & 48.32 & 12.51 & 0.77 & 0.02 & 0.04\nl
  7 &  94.04 & 48.67 & 14.70 & 0.59 & 0.01 & 0.04\nl
\multicolumn{7}{c}{...}\nl
\tablenotetext{a}{Numbers larger than 10,000 are composed of 
the numbers from Clement (\markcite{cle99}1999) plus 10,000 
for cross-identification purposes.}
\tablecomments{The complete data set is available in the electronic 
form of the Journal.}
\enddata
\end{deluxetable}

\begin{deluxetable}{ccccc}
\tablenum{3}
\tablewidth{0pt}
\tablecaption{Mean standard deviations from the artificial star 
simulations (see Section 2 for details). 
\label{tbl-3}}
\tablehead{
\multicolumn{1}{c}{Magnitude} & 
\multicolumn{2}{c}{$\rm M\,56$} & 
\multicolumn{2}{c}{$\rm M\,15$}\nl
\multicolumn{1}{c}{Bin} & 
\multicolumn{1}{c}{$\rm J$} & \multicolumn{1}{c}{$\rm K_{\rm s}$} & 
\multicolumn{1}{c}{$\rm J$} & \multicolumn{1}{c}{$\rm K_{\rm s}$}\nl
}
\startdata
10.0-12.0 & 0.04 & 0.04 & 0.05 & 0.05 \nl
12.0-14.0 & 0.05 & 0.06 & 0.07 & 0.08 \nl
14.0-16.0 & 0.10 & 0.13 & 0.15 & 0.17 \nl
\enddata
\end{deluxetable}

\begin{deluxetable}{lcccc}
\tablenum{4}
\tablewidth{0pt}
\tablecaption{Fits to the RGBs of our clusters. The slopes determined 
by Ferraro et al. (2000) are given in the last column for comparison.
\label{tbl-4}}
\tablehead{
\multicolumn{1}{c}{Cluster} & 
\multicolumn{1}{c}{r.m.s.} & 
\multicolumn{1}{c}{b} & 
\multicolumn{1}{c}{a} & 
\multicolumn{1}{c}{$\rm a_{(Ferraro~et~al.,~2000)}$} 
}
\startdata
 $\rm M\,15$    & $\pm0.04$ & $1.34\pm0.08$ & $-0.053\pm0.007$ & $-0.047\pm0.001$\nl
 $\rm M\,56$    & $\pm0.05$ & $1.32\pm0.06$ & $-0.052\pm0.005$ & - \nl
$\rm NGC\,7099$ & $\pm0.07$ & $1.07\pm0.06$ & $-0.035\pm0.004$ & $-0.043\pm0.003$\nl
$\rm NGC\,6553$ & $\pm0.06$ & $2.34\pm0.09$ & $-0.108\pm0.008$ & $-0.095\pm0.002$\nl
\tablecomments{Solutions to $\rm J-K_{\rm s}~=a\times K_{\rm s}+b$}
\enddata
\end{deluxetable}

\clearpage

\figcaption[err01.eps]{Formal DAOPHOT errors of the $\rm M\,15$ and 
$\rm M\,56$ photometry in $\rm J$ (upper panels) and $\rm K_{\rm s}$ 
(lower panels).
\label{fig-1}}

\figcaption[cmd01.eps]{Color-magnitude diagrams $\rm K_{\rm s}$ vs. 
$\rm J-K_{\rm s}$ for $\rm M\,15$ (left) and $\rm M\,56$ (right). Only 
stars with DAOPHOT errors less of than 0.06 mag for 
$\rm K_{\rm s}\leq14.0$~mag, (filled circles) and stars with errors 
less than 0.10 mag for $\rm K_{\rm s}\geq14.0$ (open circles) were 
included. See Section 3 for details. Open diamonds are variable stars. 
The typical $1\sigma$ errors are indicated on the left hand side, for 
different magnitudes. Also the effect of visible extinction of 
$\rm A_V\,=\,1.0$ mag is shown with an arrow. 
\label{fig-2}}

\figcaption[cmdf04.eps]{Color-magnitude diagrams $\rm K_{\rm s}$ vs. 
$\rm J-K_{\rm s}$ for $\rm M\,15$, $\rm M\,56$, $\rm NGC\,6553$ and 
$\rm NGC\,7099$ with the RGB linear fits. See Section 4.1 for details.
\label{fig-3}}

\figcaption[p07.eps]{RGB slope vs. $\rm[Fe/H]$ relation. The filled 
circles are $\rm M\,15$ and $\rm M\,56$, the triangles are the data from 
Kuchinski \& Frogel (1995), and the open circles are data added from 
other sources (see Section 4.1 for details). The error bars represent 
$\pm 1\sigma$ errors. The thick solid line shows our best fit and the 
thick dashed lines show $\pm 1\sigma$ error for the slope. The thin 
solid line shows the best fit and the thin dashed lines show 
$\pm 1\sigma$ error of Ferraro et al., (2000). 
\label{fig-4}}

\figcaption[c04.eps]{Relation of $\rm (J-K_{\rm s})_0(RGB)$ vs. 
$\rm[Fe/H]$. The cluster symbols are the same as in Figure 4. The thick 
solid line shows our best fit and the thick dashed lines show 
$\pm 1\sigma$ error for the slope. The thin solid line shows the best 
fit and the thin dashed lines show $1\sigma$ error of Frogel, Cohen \& 
Persson (1983). The dotted line is the fit of Minitti, Olszewski \& 
Rieke (1995). 
\label{fig-5}}

\clearpage
\plotone{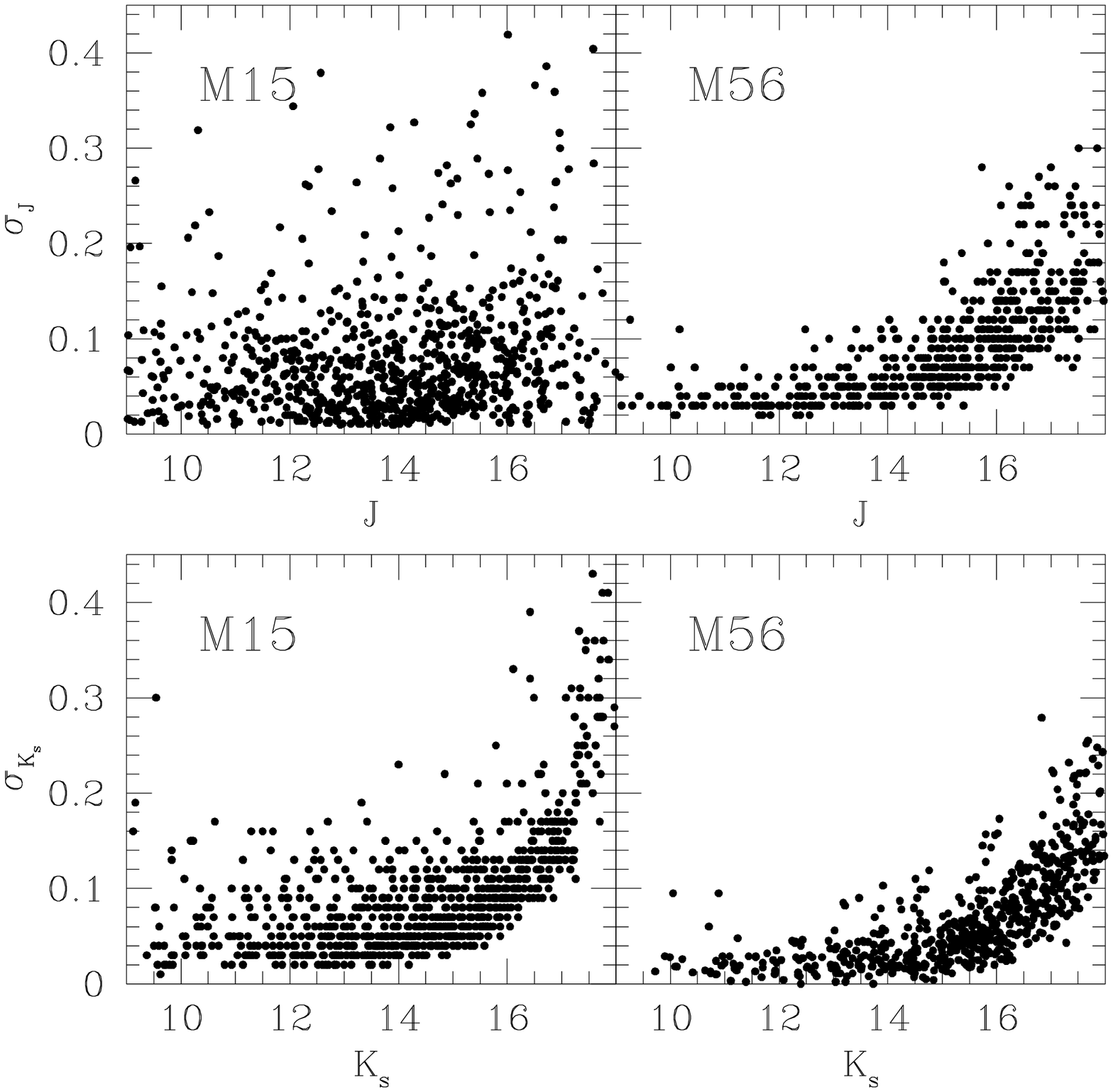}
\clearpage
\plotone{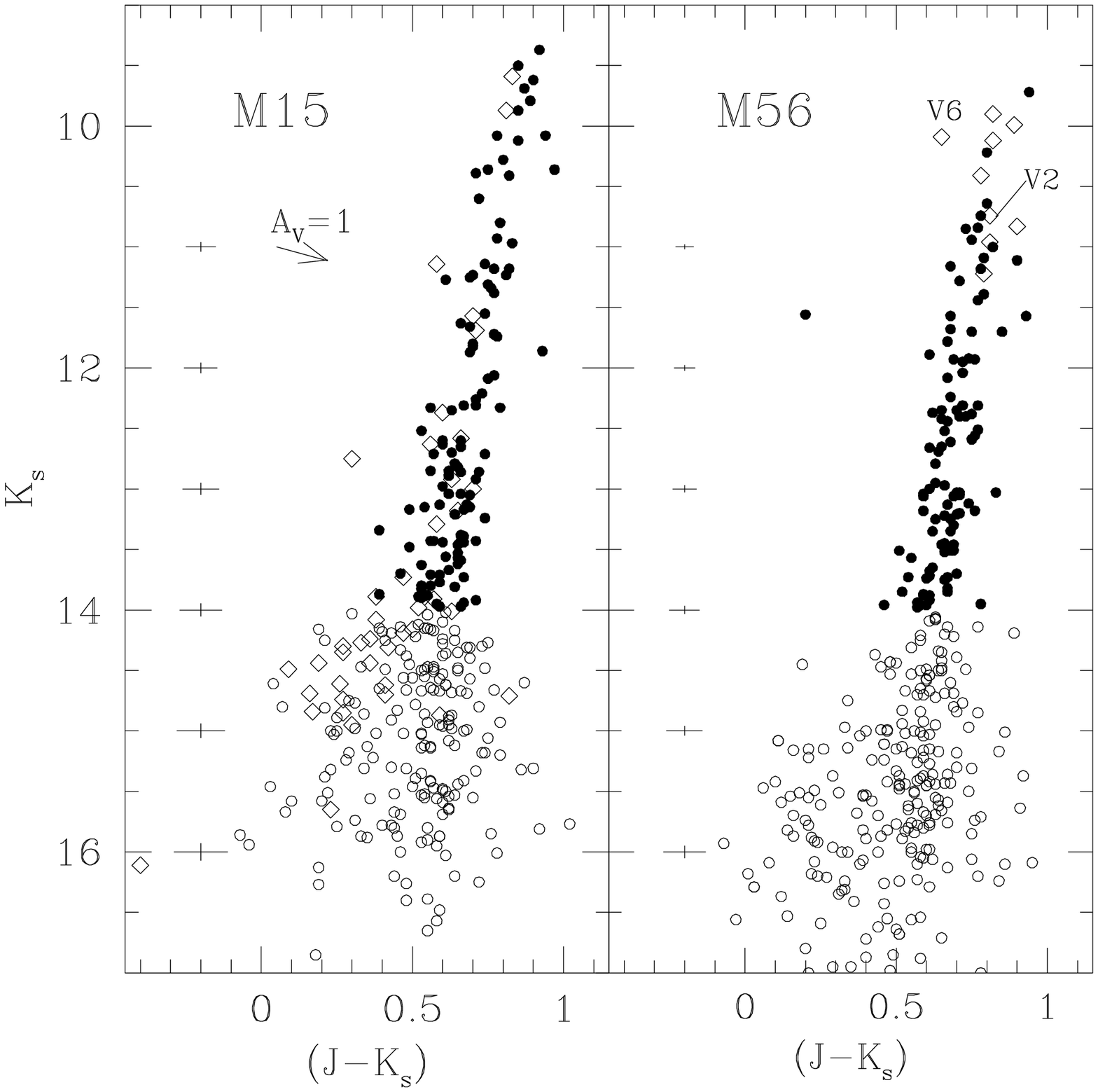}
\clearpage
\plotone{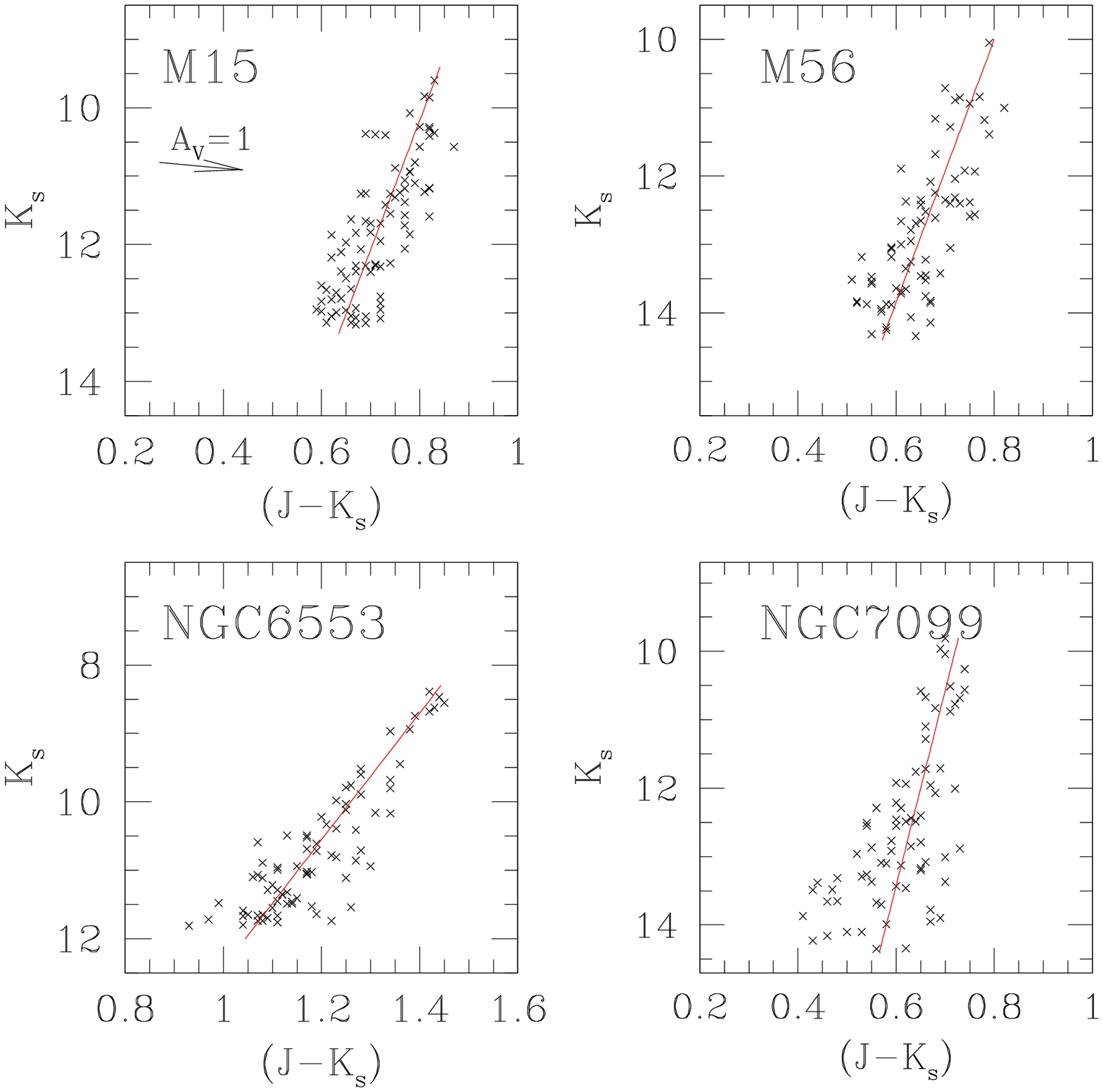}
\clearpage
\plotone{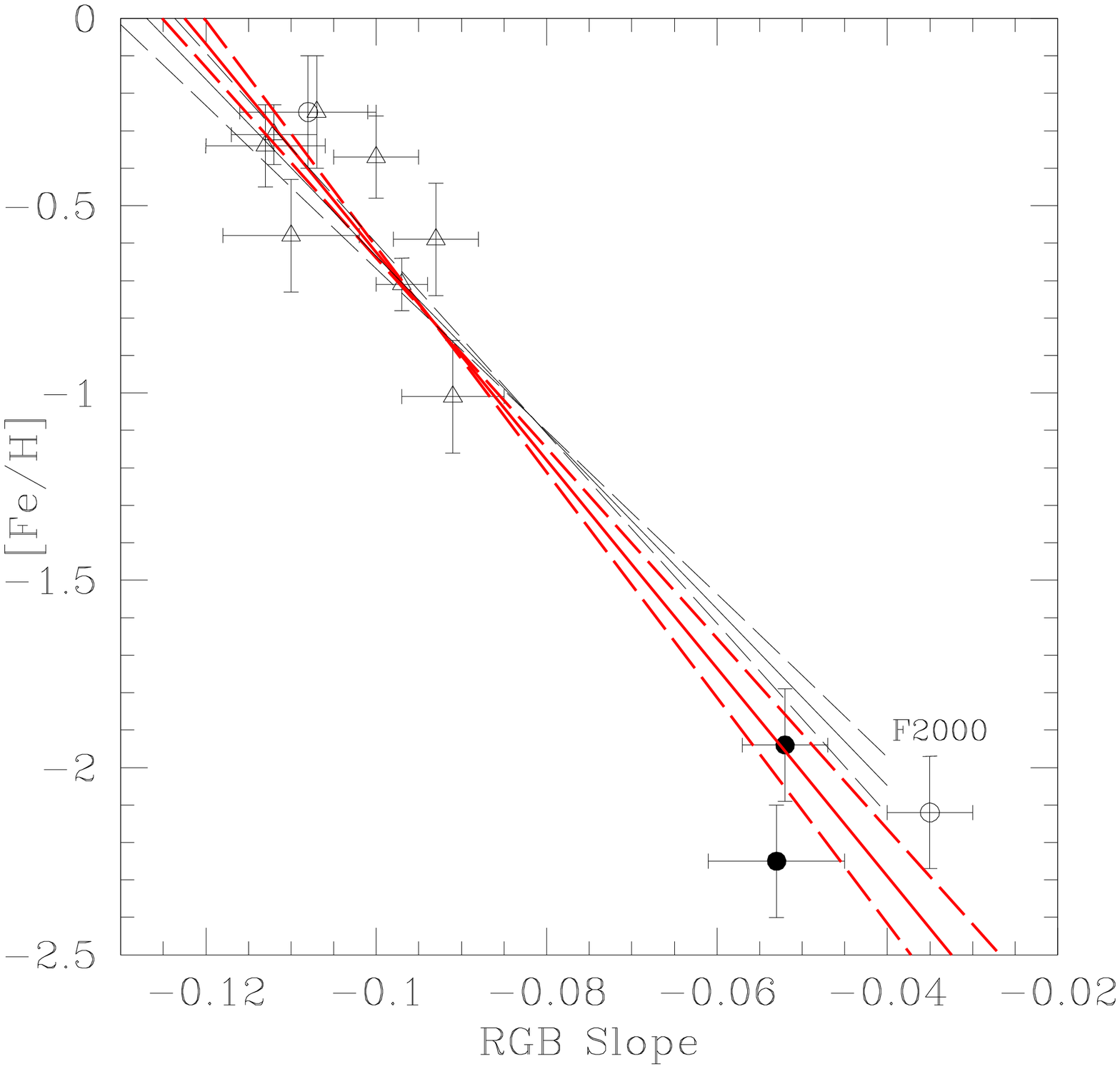}
\clearpage
\plotone{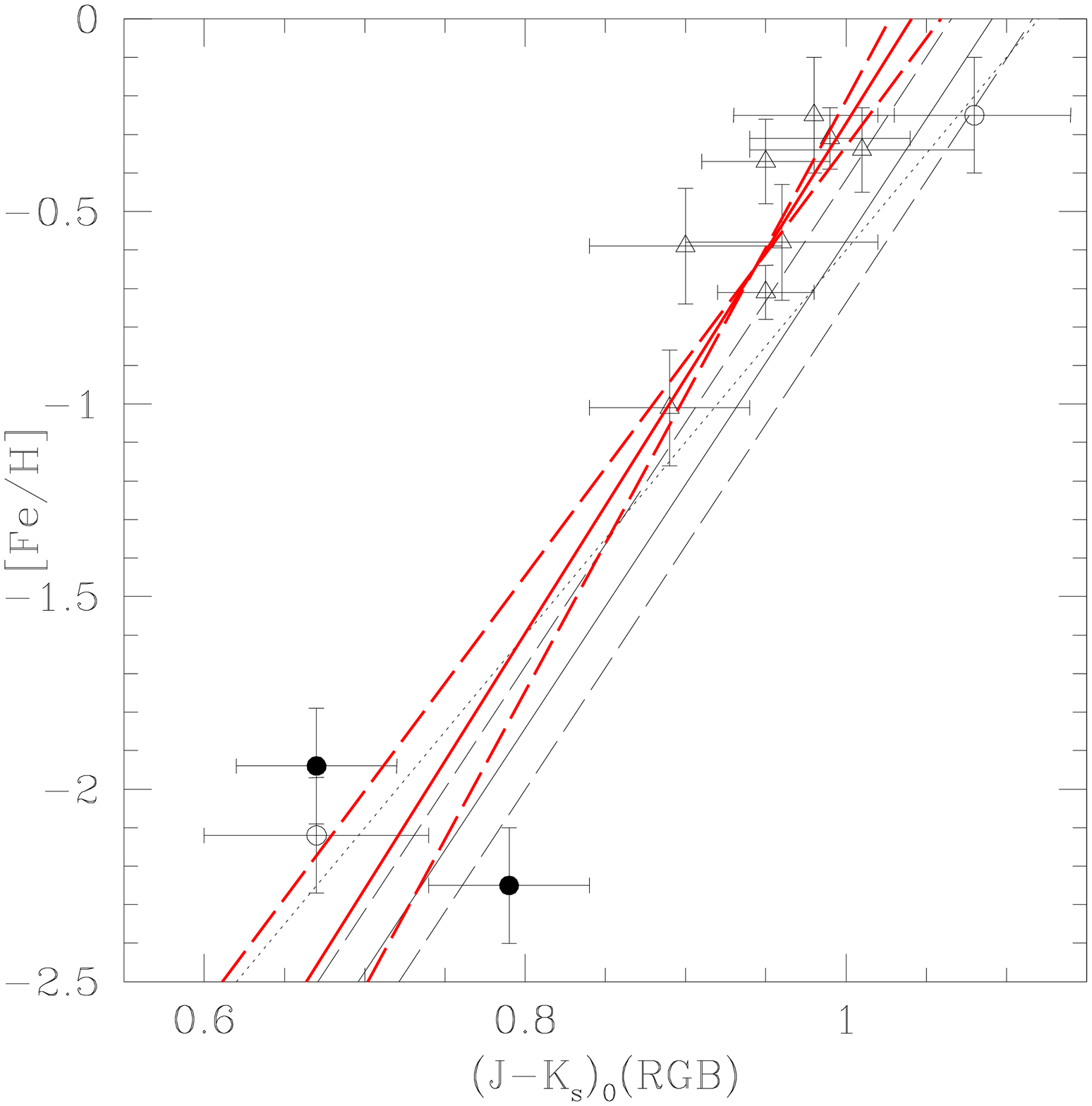}


\clearpage
\begin{references}
\reference{arm88} Armandroff, T.E., \& Zinn, R., 1988, \aj, 96, 92
\reference{aru81} Auri\`{e}re, M., \& Cordoni, J.-P., 1981, \aaps, 46, 347
\reference{bah75} Bahcall, J.N., \& Ostriker, J.P., 1975, \nat, 256, 23
\reference{bai88} Bailyn, C.D., Grindlay, J.E., Cohn, H., \& Lugger, P.M., 
	1988, \apj, 331, 303
\reference{bar65} Barbon, R., 1965, Contr. Oss. Astrof. Padova in Asiago,
	175, 63 
\reference{bra98} Braun, K., Chiboucas, K., Minske, J., \& Salgado, J, 1998, 
	\pasp, 110, 810
\reference{buo85} Buonano, R., Corsi, C.E., \& Fusi Pecci, F., 1985, 
	\aap, 145, 97
\reference{but98} Butler, R.F., Shearer, A., Redfern, R.M., Colhoun, M.,
	O'Kane, P., Penny, A.J., Morris, P.W., Griffiths, W.K., \& 
	Cullum, M., 1998, \mnras, 296, 379
\reference{car97} Caretta, E., \& Gratton, R., 1997, \aaps, 121, 95
\reference{car92} Carney, B.W., Storm, J., \& Johnes, R.Y., 1992, \apj, 
	386, 663
\reference{ced92} Cederbloom, S.E., Moss, M.J., Cohn, H.N., Lugger, P.M., 
	Bailyn, C.D., Grindlay, J.E., McClure, R.D., 1992, \aj, 103, 480
\reference{cle99} Clement, C.M., 1999, "An update to Helen Sawyer Hogg's 
	third catalog of variable stars in globular cluster", in preparation
\reference{coh95} Cohen, J.G., \& Sleepers, C., 1995, \aj, 109, 242
\reference{dur93} Durrell, D., \& Harris, W., 1993, \aj, 105, 1420
\reference{eli82} Elias, J.H., Frogel, J.A., Matthews, K., \& Neugebauer, 
	G. 1982, \aj, 87, 1029
\reference{fer93} Ferraro, F.R., \& Parsce, F., 1993, \aj, 106, 154
\reference{fer00} Ferraro, F.R., Montegriffo, P., Origlia, L., \& Fusi Pecci, 
	F., 2000, preprint (astro-ph/9912265)
\reference{fro83} Frogel, J.A., Cohen , J.G., \& Persson, S.E., 1983, 
	\apj, 275, 773
\reference{fro88} Frogel, J., \& Elias, J., 1988, \apj, 324, 823
\reference{gru99} Grundahl, F., Catelan, M., Landsman, W.B., Stetson, 
	P.B., \& Andersen, M.I., 1999, \apj, 524, 242
\reference{gua98} Guarnieri, M.D., Ortolani, S., Montegriffo, P., Renzini, 
	A., Barbuy, B., Bica, E., \& Moneti, A., 1998, \aap, 331, 70
\reference{har79} Harris, W.E., \& Racine, R., 1979, \araa, 17, 241
\reference{har96} Harris, W.E., 1996, \aj, 112, 1487
\reference{kin75} King, I.R., 1975, in Dynamics of Stellar Systems, Proc.,
	IAU Symposium No. 69, edited by A. Hayli (Reidel, Dordrecht), p.69
\reference{kuc95} Kuchinski, L.E., Frogel, J.A., \& Terndrup D.M., 1995, 
	\aj, 109, 1131
\reference{kuc95b} Kuchinski, L., \& Frogel, J., 1995, AJ, 110, 2844
\reference{min95} Minitti, D., Olszewski, E.W., \& Rieke, M., 1995, \aj, 
	110, 1686
\reference{mon95} Montegriffo, P., Ferraro, F., Fusi Pecci, F., \&
	Origlia, L., 1995, \mnras, 276, 739
\reference{per98} Persson, S.E., Murphy, D.C., Krzeminski, W., Roth, M., 
	\& Rieke, M.J., 1998, \aj, 116, 2475
\reference{rie85} Rieke, G.H., \& Lebofsky, M.J., 1985, \apj, 288, 618
\reference{rus81} Rishel, B., Sanders, W., \& Schroder, R., 1981, \aap, 
	45, 443
\reference{ros51} Rosino, L., 1951, Co. Asiago, No. 21, 135
\reference{ros61} Rosino, L., 1961, Co. Asiago, No. 117
\reference{rus99} Russeva, T., 2000, in preparation  
\reference{san70} Sandage, A.R., 1970, \apj, 162, 841
\reference{saw40} Sawyer, H.B., 1940, Publ. David Dunlop Obs., 1, No. 5
\reference{saw49} Sawyer, H.B., 1949, J.R.A.S.C., 43, 38
\reference{smr95} Smriglio, F., Dasgupta, A.K., \& Boyle, R.P., 1995, 
	Balt. Astr., 5, 451
\reference{ste93} Stetson, P., 1993, User's Manual for DAOPHOT II
\reference{tie97} Tiede, G., Martini, P., \& Frogel, J., 1997, \aj, 114, 694
\reference{weh85} Wehlau, A., \& Sawyer Hogg, H., 1985, \aj, 90, 2514
\reference{yan94} Yanny, B., Guhathakurta, P., Bahcall, J.N., \& Schneider,
	D.P., 1994, \aj, 107, 1745

\end{references}
\end{document}